\documentclass[12pt]{article}
\usepackage{amsmath,amssymb,amsfonts,amsthm}

\title{Barut-Girardello and Gilmore-Perelomov coherent states for pseudoharmonic
oscillator and their nonclassical properties: factorization method}

\author{{ M K Tavassoly$^{1,2,3,*}$ and H R Jalali$^1$} \\
 \footnotesize{$^1$ Atomic and Molecular Group, Faculty of Physics, Yazd University, Yazd, Iran} \\
 \footnotesize{$^2$ Photonics Research Group, Engineering Research Center, Yazd University, Yazd, Iran} \\
 \footnotesize{$^3$ The Laboratory of Quantum Information Processing, Yazd University, Yazd, Iran}\\
 \footnotesize{$^*$ E-mail: mktavassoly@yazd.ac.ir}}

\begin{document}

\maketitle


%
%
\begin{abstract}
In this paper we try to introduce the ladder operators associated with the pseudoharmonic oscillator, after solving the corresponding Schr\"{o}dinger equation by using the factorization method. The obtained generalized raising and lowering operators naturally lead us to the Dirac representation space of the system which is very easier to work with, in comparison to the functional Hilbert space.  The $SU(1,1)$ dynamical symmetry group associated with the considered system is exactly established through investigating  the fact that the deduced operators satisfy appropriate commutation relations.
This result enables us to construct two important and distinct classes of Barut-Girardello and Gilmore-Perelomov coherent states associated with the system. Finally, their identities as the most important task are exactly resolved  and some of their nonclassical properties are illustrated, numerically.
 \end{abstract}


  \section{Introduction}\label{sec-intro}
   Molecular vibrational and rotational spectroscopies are usually known as the important parts of molecular physics. The harmonic oscillator could be considered as a useful ground to explain the  molecular vibrations, but this model is very ideal and simple and so has some particular restrictions. Strictly speaking, the molecular vibrations are really anharmonic, so if their Schr\"{o}dinger equation can be  exactly solved,
   they may provide more suitable models for the description of vibrating molecules. As some examples of anharmonic oscillator molecular potentials we may refer to the Morse \cite{morse}, P\"{o}schl-Teller \cite{teller1,teller2} and Mie-type \cite{mie} potentials which are three typical model potentials which the molecular vibrations can be described satisfactorily by means of them. In this work, we want to study another anharmonic potential, the so-called pseudoharmonic oscillator,
   which is the sum of the harmonic oscillator and the inversely quadratic potential \cite{pseudo}, i.e.,
 \begin{equation}\label{pseudoharmonic}
 V(x)=\frac{1}{2}m\omega^{2}x^{2}+\frac{\hbar^{2}}{2m}\frac{\alpha^{2}}{x^{2}},
 \end{equation}
   where $m$, $\omega$ and $\alpha$ respectively represent the mass of the particle, the frequency and the strength of the external field.
  Sometimes this system has been called isotonic potential \cite{isotonic1, isotonic2, isotonic3, isotonic4}.
  In fact, this potential was proposed by Goldman and Krivchenkov \cite{goldman}.  Some methods that have been used to solve the wave equation for the pseudoharmonic potential can be summarized  as follows: Nikiforov-Uvarov method \cite{nikiforov1, nikiforov2}, polynomial solution \cite{polynomial}, exact quantization rule \cite{quantization} and  Laplace transform approach \cite{laplace}.
  Moreover, obtaining the exact solutions of the Schr\"{o}dinger equation for the pseudoharmonic oscillator using a new method
  (and in general enlarging the circle of solvable quantum systems) is one of the main problems in quantum physics \cite{zare}.

  In the present study we firstly want to solve the latter equation by the factorization method which simply arrives us to the exact solution of the eigenvalues and eigenfunctions of the considered system. Very recently this method is used for the construction of  a new set of generalized coherent state and then the geometric phase of the states is investigated \cite{Fernandz} (see  also \cite{Fernandz1, Fernandz2, Fernandz3, Fernandz4} and references therein).
  Our motivation lies in the fact that, by using this method we will be able to find out the appropriate generalized creation and annihilation operators for the chosen system, by which we are simply lead to the Dirac space with its instruments (ket-bra space) provides us with a very simpler tool than the usual functional space.
  Consequently, we can construct two new, important and distinct classes of generalized coherent states associated with such physical system namely Barut-Girardello and Gilmore-Perelomov coherent states.
  After this exploration, due to the importance of the nonclassicality of quantum states in various theoretical and experimental fields of physics in recent decades
  \cite{cpb11, cpb33} we are naturally lead to the investigation of the nonclassically signs of our introduced states for completing the work. Moreover, prior to these examinations we will exactly establish the resolution of the identity of both classes of states, as the most important property of any generalized coherent state.

  Notice that our present work can be considered in the line of the dynamical generalization of coherent states \cite{Nieto1, Nieto2, Nieto3}. Also, it is worth mentioning that, the obtained Barut-Girardello and Gilmore-Perelomov coherent states also can be really categorized in the "algebraic" \cite{Manko, RokniTav1, RokniTav2} and "group theoretical" \cite{GP1, GP2, BG} generalization of coherent states, respectively.
  From the physical point of view, we would like to mention that the strong relation of the laser light to the well-known standard bosonic coherent states corresponds to harmonic oscillator is well established from the sixth decades ego by the pioneering works of Glauber, Sudarshan and Klauder. Indeed, this was only after this discovery that, hard works along their generalizations as well as physical realization were began.
  This is why their various manners of generalizations are yet of hot importance \cite{cpb1, cpb2,  cpb6,  cpb3, cpb4}.
  Henceforth, apart from presenting a powerful technique which can be used for other known potentials, due to the direct relation of our introduced coherent states to a particular  physical potential (which as we mentioned in the above lines, may be more suitable for the description of real physical phenomena) we hope that our work suitably concerns to physical realization, too.

As is mentioned before,  this potential has received  a lot of attention in recent years.
In addition, the used approach in this paper is conceptually near to the well-known SUSY  quantum mechanics techniques
which is an interesting approach for many authors \cite{newrefs1, newrefs2, newrefs3, newrefs4, newrefs5, newrefs6, newrefs7}.
Altogether, it may clearly be seen that our method to the solution of the eigenvalue problem
and the from the final obtained results are essentially new. In this way, to the best of our knowledge, the approach to the treatment of this particular problem, the corresponding ladder operators, the associated coherent states of the types which are obtained,
in addition to establishing the resolution of the identity for both classes of states and the related nonclassicality signs have not been appeared in the literature
up to now.

  The organization of the paper is as follows. In the next Section, we are going to review the factorization method briefly.  This method is used to solve the Schr\"{o}dinger equation for pseudoharmonic oscillator in one-dimensional space in Section 3. According to our obtained results, we construct the creation and annihilation operators, by which we study the dynamical group corresponding to this system in Section 4. Then, after producing the Barut-Girardello and Gilmore-Perelomov coherent states associated with the pseudoharmonic potential, the resolution of the identity corresponding to both classes of introduced states is established in Section 5. We demonstrate some of the nonclassical properties of the states like squeezing and sub-Poissonian statistics, numerically in Section 6. Finally,   summary and conclusion are presented in Section 7.

 \section{The factorization method: a brief review}\label{sec-nl}
  The factorization method is a useful technique that reduces the dynamical equation of a given system into a simple one which is easier to handle.
  This method was first introduced by Schr\"{o}dinger \cite{schrodinger1,schrodinger2,schrodinger3} and then followed by Dirac \cite{dirac} to avoid the usage of cumbersome mathematical tools and it has also been successfully applied to resolve some problems for which there exists exact solutions.
   This method further developed by Infeld and Hull \cite{hull} and Green \cite{green}. The characteristics and advantages of the factorization method can be summarized as follows. First, this method applies only to the discrete energy spectra. Secondly, we can achieve immediately the desired eigenvalues and the normalized eigenfunctions from the given Hamiltonian without any need to use the traditional methods  in quantum mechanics. Thirdly, it is possible to avoid deriving the normalization constant, which is sometimes difficult to obtain. The fourth property comes from the fact that we may discover the hidden symmetry of the quantum system through the establishment of an appropriate Lie algebra, which usually can be realized by means of the ladder operators. This method is a very powerful tool for solving second-order differential equations and has attracted much attention of many authors. In this method, which is closely related to SUSY quantum mechanics,
   the second-order differential operator $H_0$ (Hamiltonian) is written as the product of two first-order differential operators $a_{0}$ and $a_{0}^{\dagger}$ plus a constant $E_{0}$,
 \begin{equation}\label{susym1}
   H_0=a_0^\dag a_0 +E_0.
 \end{equation}
  Then, rewriting the SUSY Hamiltonians  $H_-$ and $H_+$ respectively as  $H_0$ and $H_1$, and
  changing suitably the subscripts we have  $H_1=a_0 a_0^\dag +E_0$. Also, by factorizing $H_1$ one has
  $H_1=a_1^\dag a_1 +E_1$, so firstly  we have
 \begin{equation}\label{susym2}
   H_1=a_0 a_0^\dag +E_0= a_1^\dag a_1 +E_1,
 \end{equation}
  with the SUSY partner of $H_1$ as $H_2=a_1 a_1^\dag  + E_1$.
  Continuing the above procedure leads to the
  generation of a hierarchy of Hamiltonians \cite{sukumar1,sukumar2} given by
  \begin{equation}\label{recurrence}
   H_{n+1}= a_{n+1}^{\dagger}a_{n+1}+E_{n+1}=a_{n}a_{n}^{\dagger}+E_{n},  \;\; n=0,1,2,3,...\;.
 \end{equation}
    The complete set of eigenfunctions can be generated by the successive action of these operators on the key eigenfunctions, which are the exact solutions of the first-order differential equation. Then, the recurrence relation may be used to achieve all eigenvalues.\\
  In fact, with the following theorem we can find eigenfunctions and eigenvalues \cite{ohanian}.\\
 \textbf{Theorem}: Assume that equations (\ref{susym1}) to (\ref{recurrence})  with real constants $E_{n}$ are true. Also suppose that there exists a null eigenfunction $|\xi_{n}\rangle$ with zero eigenvalue for each $a_{n}$, i.e.,
 \begin{equation}\label{Help}
   a_{n}|\xi_{n}\rangle=0.
 \end{equation}
 Then,

 a) $E_{n}$ is the $n$th eigenvalue of the Hamiltonian.

 b) The corresponding eigenfunction of the Hamiltonian $H$ is,
 \begin{equation}\label{eigenfunction}
   |\psi_{n}\rangle=c_{n}a_{0}^{\dagger}a_{1}^{\dagger}...a_{n-1}^{\dagger}|\xi_{n}\rangle,
 \end{equation}
 where $c_{n}$ is a normalization constant may be determined as:
 \begin{equation}\label{normalization}
   |c_{n}|=[(E_{n}-E_{n-1})(E_{n}-E_{n-2})...(E_{n}-E_{0})]^{-1/2}.
 \end{equation}
Three important worthwhile points lie in the theorem:

i) This method applies only to the discrete energy spectra.

ii) In the selection of $a_n$, one should choose it in such a way that $E_{n}$ gets the maximum value.

iii) The theorem doesn't give us any information about the form of $a_{n}$. Moreover, evidently the form of $a_{n}$ depends explicitly
 on the particular potential $V(x)$. Anyway, since $a_{0}^{\dagger}a_{0}$ is equal to $H$ except for a constant, thus $a_{0}$ must contain a linear momentum term to be consistent with the kinetic energy part of the Hamiltonian. Accordingly, $a_{n}$ is assumed to have the form
 \begin{equation}\label{annihil-opera}
   a_{n}=\frac{1}{\sqrt{2m}}(p+if_{n}(x)),
 \end{equation}
 where $p$ is the momentum operator and $f_{n}(x)$ is a real (and differentiable) unknown function of $x$. This function would be such that equation (\ref{recurrence}) is satisfied.

\section{Solving the Schr\"{o}dinger equation of pseudoharmonic oscillator: factorization method}\label{sec-n2}
Now, we are going to search for the eigenenergies and eigenfunctions of a particular system, i.e. pseudoharmonic potential in one-dimensional space, by the factorization method. Let us start with the Schr\"{o}dinger equation for this potential, i.e.,
 \begin{equation}\label{schrodinger}
 \left(\frac{p^{2}}{2m}+V(x)\right)\psi(x)=E\psi(x),
 \end{equation}
 where the operator $p$ is defined as $\frac{\hbar}{i}\frac{d}{dx}$.
 According to the form of pseudoharmonic oscillator potential introduced in (\ref{pseudoharmonic}) and keeping in mind (\ref{annihil-opera}) we can guess that $f_{n}(x)$ can take the following form
\begin{equation}\label{fn}
 f_{n}(x)=b_{n}x + \frac{c_{n}}{x},
 \end{equation}
where $b_{n}$ and $c_{n}$ are constant values which should be chosen  such that $E_{n}$ gets the maximum value. Thus, $a_{n}$ can be written as:
\begin{equation}\label{an bn cn}
 a_{n}=\frac{1}{\sqrt{2m}} \left[p+i\left(b_{n}x+\frac{c_{n}}{x}\right)\right].
 \end{equation}
With the help of $a_{n}$ and its conjugate $a_{n}^{\dagger}$, one can easily check that the following relations hold:
\begin{equation}\label{an dag an}
 a_{n}^{\dagger}a_{n}=\frac{1}{2m}\left[p^{2}+b_{n}^{2}x^{2}+b_{n}\left(2c_{n}+\hbar\right)+\frac{c_{n}}{x^{2}}\left(c_{n}-\hbar\right)\right],
 \end{equation}
\begin{equation}\label{an an dag}
 a_{n}a_{n}^{\dagger}=\frac{1}{2m}\left[p^{2}+b_{n}^{2}x^{2}+b_{n}\left(2c_{n}-\hbar\right)+\frac{c_{n}}{x^{2}}\left(c_{n}+\hbar\right)\right].
 \end{equation}
 With substituting $n=0$ in (\ref{an dag an}) and using (\ref{recurrence}) it is  required that
 \begin{eqnarray}\label{b0 c0}
 \frac{1}{2m}\left[p^{2}+b_{0}^{2}x^{2}+b_{0}\left(2c_{0}+\hbar\right)+\frac{c_{0}}{x^{2}}\left(c_{0}-\hbar\right)\right]+E_{0} \nonumber\\
 =\frac{p^{2}}{2m}+\frac{1}{2}m\omega^{2}x^{2}+\frac{\hbar^{2}}{2m}\frac{\alpha}{x^{2}}.
 \end{eqnarray}
The above equation will be satisfied, if
 \begin{equation}\label{b0}
b_{0}^{2}=m^{2}\omega^{2}\;\;\;\;\longrightarrow\;\;\;\;b_{0}=-m\omega,
 \end{equation}
 \begin{equation}\label{c0}
c_{0}(c_{0}-\hbar)=s(s+1)\hbar^{2}\;\;\;\;\longrightarrow\;\;\;\;c_{0}=(s+1)\hbar,
 \end{equation}
 \begin{equation}\label{b0 c0 E0}
b_{0}(2c_{0}+\hbar)+2mE_{0}=0,
 \end{equation}
 where the allowed values of $b_0$ in (\ref{b0})  can satisfy the condition (ii)  of the stated theorem for which $E_0$ gets the maximum value. Also, note that the change of variable $s=-\frac{1}{2}+\sqrt{\alpha+\frac{1}{4}}$ in (\ref{c0}) has been used.
 With substituting (\ref{b0}) and (\ref{c0}) in (\ref{b0 c0 E0}), the maximum of the ground state energy will be achieved as:
 \begin{equation}\label{E0}
 E_{0}=\hbar\omega\left(s+\frac{3}{2}\right).
 \end{equation}
 For attaining the other eigenvalues we can use the recurrence relation (\ref{recurrence}), from which one has:
 \begin{eqnarray}\label{bn+1 cn+1}
 \frac{1}{2m}\left[p^{2}+b_{n+1}^{2}x^{2}+b_{n+1}\left(2c_{n+1}+\hbar\right)+\frac{c_{n+1}}{x^{2}}\left(c_{n+1}-\hbar\right)\right]+E_{n+1} \nonumber\\
 =\frac{1}{2m}\left[p^{2}+b_{n}^{2}x^{2}+b_{n}\left(2c_{n}-\hbar\right)+\frac{c_{n}}{x^{2}}\left(c_{n}+\hbar\right)\right]+E_{n}.
 \end{eqnarray}
 The relation (\ref{bn+1 cn+1}) will be satisfied, if
 \begin{equation}\label{bn}
b_{n+1}^{2}=b_{n}^{2}\;\;\;\;\longrightarrow\;\;\;\;b_{n}=b_{0}=-m\omega,
 \end{equation}
 \begin{equation}\label{cn}
c_{n+1}(c_{n+1}-\hbar)=c_{n}(c_{n}+\hbar)\;\;\;\;\longrightarrow\;\;\;\;c_{n}=(s+n+1)\hbar,
 \end{equation}
 \begin{equation}\label{bn cn En}
b_{n+1}(2c_{n+1}+\hbar)+2mE_{n+1}=b_{n}(2c_{n}-\hbar)+2mE_{n}.
 \end{equation}
Finally, from (\ref{bn}), (\ref{cn}) and (\ref{bn cn En}) we can derive the required eigenvalues as
\begin{equation}\label{En}
 E_{n}=\hbar\omega\left(2n+s+\frac{3}{2}\right),\;\;\;n=0,1,2,...\; .
 \end{equation}
 With substituting $b_{n}$ and $c_{n}$, respectively from (\ref{bn}) and (\ref{cn}) in (\ref{an bn cn}), $a_{n}(x)$ will be explicitly obtained
 \begin{equation}\label{anx}
 a_{n}(x)=\frac{1}{\sqrt{2m}} \left[p+i\left(-m\omega x+\frac{(s+n+1)\hbar}{x}\right)\right].
 \end{equation}
The equation (\ref{En}) gives all eigenvalues, if there exists an appropriate answer for any arbitrary $n$ in (\ref{Help}). Thus, we need to solve the following equation
 \begin{equation}\label{ann}
 a_{n}(x) \xi_{n}(x)=0
 \end{equation}
 or equivalently
 \begin{equation}\label{solutionan}
 \frac{d\xi_{n}(x)}{dx}=\left(\frac{s+n+1}{x}-\frac{m\omega}{\hbar}x \right)\xi_{n}(x).
 \end{equation}
 The solution of the above differential equation can be simply obtained as follows
 \begin{equation}\label{xi}
 \xi_{n}(x)=x^{s+n+1}e^{-\frac{m\omega}{2\hbar}x^{2}}.
 \end{equation}
 By using (\ref{eigenfunction}) and after doing some tedious calculations, the normalized eigenfunctions of pseudoharmonic oscillator can be obtained as ($\hbar=m=\omega=1$):
\begin{equation}\label{psi}
 \psi_{n}(x)=N_{n}x^{s+1}e^{-\frac{x^{2}}{2}}L_{n}^{s+\frac{1}{2}}(x^{2}),
 \end{equation}
 where $L_{n}^{s+\frac{1}{2}}(x^{2})$ is the associated Laguerre polynomials and the normalization constant in (\ref{psi}) is determined as
 \begin{equation}\label{Nn}
N_{n}=\sqrt{\frac{2\Gamma(n+1)}{\Gamma(n+s+\frac{3}{2})}}.
 \end{equation}

 \section{Introducing ladder operators associated with pseudoharmonic oscillator} \label{sec-n3}

 It is remarkable that the operators $a_n$ in (\ref{anx}) and its conjugate do not play the role of raising
 and lowering operators for the eigenstates of the pseudoharmonic potential. But, keeping in mind these operators, we want to construct
 a rather new pair of operators which yield our mentioned requirement. For this purpose, we introduce a new operator as follows
\begin{equation}\label{An}
   A_{n}\equiv-\frac{d}{dx}-x+\frac{s+n+1}{x},
 \end{equation}
 where its conjugate read as
 \begin{equation}\label{And}
    A_{n}^{\dagger}\equiv\frac{d}{dx}-x+\frac{s+n+1}{x}.
 \end{equation}
 By applying the operator $A_{n}$ on $\psi_{n}(x)$ and using the following recurrence relations for the associated Laguerre polynomials \cite{Laguerre}
 \begin{equation}\label{Laguerre}
    x\frac{d}{dx}L_{n}^{\alpha}(x)=\left\{
                                  \begin{array}{ll}
                                    nL_{n}^{\alpha}(x)-(n+\alpha)L_{n-1}^{\alpha}(x), &  \\
                                    (n+1)L_{n+1}^{\alpha}(x)-(n+\alpha+1-x)L_{n}^{\alpha}(x),
                                  \end{array}
                                \right.
 \end{equation}
   we have
 \begin{equation}\label{Anpsi}
   A_n\psi_{n}(x)=-\frac{n}{x}\psi_{n}(x)+\frac{2}{x}\sqrt{n(n+s+\frac{1}{2})}\; \psi_{n-1}(x).
 \end{equation}
 Similarly, by acting $A_n^{\dagger}$ on $\psi_{n}(x)$ one has:
 \begin{equation}\label{Andpsi}
   A_n^{\dagger}\psi_{n}(x)=-\frac{n+1}{x}\psi_{n}(x)+\frac{2}{x}\sqrt{(n+1)(n+s+\frac{3}{2})}\; \psi_{n+1}(x).
 \end{equation}
 Clearly, the above operators $A_{n}, A_{n}^{\dagger}$ do not yet possess the role of ladder operators. Altogether,
 it is possible to find out the ladder operators which can regularly raise and lower any eigenfunctions of our considered system, i.e.,
\begin{equation}\label{Mmp}
   M_{\pm}=\pm\frac{x}{2}\frac{d}{dx}+N+\frac{s}{2}+\frac{1}{\frac{3}{2}\mp\frac{1}{2}}-\frac{x^{2}}{2},
 \end{equation}
 where we have introduced the number operator $N$ which satisfies the eigenvalue equation
 \begin{equation}\label{num oper}
 N|n\rangle = n |n\rangle,\;\;\;\;\;\psi_{n}(x)= \langle x|n\rangle
 \end{equation}
 with $\{|n\rangle\}_{n=0}^\infty$ as the eigenkets of the pseudoharmonic oscillator system.
 Notice that $N \neq M_+ M_- $. It is easy to check that
  the ladder operators $M_{\pm}$ have the following properties
 \begin{equation}\label{Mmp n}
  M_{\pm}|n\rangle=m_{\pm}|n\pm 1\rangle,\;\;\;\;\;n=0,1,2,...
 \end{equation}
 where
 \begin{equation}\label{mkmp}
  m_{\pm}=\sqrt{(n+\frac{1}{2} \pm \frac{1}{2})(n+s+\frac{2\pm 1}{2})}.
 \end{equation}
 We now study the dynamical group associated with the operators $M_+$ and $M_-$. Based on the obtained results in (\ref{Mmp n}) and (\ref{mkmp}), we can calculate the following commutator:
 \begin{equation}\label{Mm Mp}
  [M_-,M_+]|n\rangle=2m_0 |n\rangle,
 \end{equation}
 where we have defined
 \begin{equation}\label{mk0}
  m_0=n+\frac{s}{2}+\frac{3}{4}.
 \end{equation}
 Thus, we can define the operator
 \begin{equation}\label{M0}
  M_0=N+\frac{s}{2}+\frac{3}{4}.
 \end{equation}
 Adding our above results, the operators $M_\pm$ and $M_0$ justify the following algebraic  relations
  \begin{equation}\label{su}
  [M_-,M_+]=2M_0,\;\;\;\;\;[M_0,M_\pm]=\pm M_\pm.
 \end{equation}
 As is well known, according to (\ref{su}), the operators $M_\pm$ and $M_0$ correspond to an $su(1,1)$ Lie algebra. In addition, the system Hamiltonian can be simply expressed as:
 \begin{equation}\label{H}
  H |n \rangle=(2n+s+\frac{3}{2}) |n \rangle =[M_-,M_+] |n \rangle.
 \end{equation}
 The form of the Hamiltonian in (\ref{H}) satisfy the eigenvalue equation $H|n\rangle=E_n |n\rangle$ with $E_n$ determined in (\ref{En}).
 It may be interesting to notice that while for the harmonic oscillator one has $H_{HO}|n\rangle_{HO}= (n+1/2)|n\rangle_{HO}= \{a, a^\dag\}|n\rangle_{HO}$, where $a$, $a^{\dagger}$ are the standard bosonic annihilation and creation operators,
 the anti-commutation relation for harmonic oscillator gives its place to commutation relation for pseudoharmonic oscillator.\\
 %

   \section{Coherent states for pseudoharmonic oscillator: Barut-Girardello  and Gilmore-Perelomov }\label{sec-n4}
   We are now able to produce two distinct classes of coherent states for the considered system.
Usually, the Barut-Girardello type of coherent state is defined as the right eigenstate of the annihilation operator, i.e.,
\begin{equation}\label{Lm alpha}
   M_{-}|z \rangle_{BG}=z |z \rangle_{BG}.
 \end{equation}
 To calculate the explicit form of  $|z \rangle_{BG}$ with the traditional
 method, we write
 \begin{equation}\label{alpha is sigma}
   |z \rangle_{BG}=\sum_{n=0}^{\infty} c_{n}|n\rangle.
 \end{equation}
 Setting it in (\ref{Lm alpha}) and making use of (\ref{Mmp n}) finally arrives us at the explicit form of coherent state
 \begin{equation}\label{Barut}
   |z \rangle_{BG}=N_{BG}(|z|^{2})\sum_{n=0}^{\infty}
   \left[\frac{\Gamma(s+\frac{3}{2})}{\Gamma(n+1)\Gamma(n+s+\frac{3}{2})}\right]^{1/2}z^{n}|n\rangle,
 \end{equation}
 where the factor $N_{BG}(|z|^{2})$ can be calculated by the
 normalization condition: $\langle z |z \rangle =1$, i.e.
 \begin{equation}\label{normal Barut}
   N_{BG}(|z|^{2})=\left({_{0}F_{1}(s+\frac{3}{2},|z|^{2})}\right)^{-\frac 1 2}.
 \end{equation}
 where $_{0}F_{1}(a,x)$ denotes the hypergeometric function.

  On the other hand, referring the $su(1,1)$ symmetry algebra of the generators as in (\ref{su}), we are able to construct the  Gilmore-Perelomov coherent state associated with the pseudoharmonic potential as follows \cite{Miri_Tavassoly_Scripta}:
  \begin{eqnarray}\label{displace oper}
   |z \rangle_{GP}&=&D(z)|0\rangle=\exp(\xi M_{+}-\xi^{\ast} M_{-})|0\rangle \nonumber \\
   &=&\exp(z M_{+})(1-|z|^{2})^{M_{0}} \exp(-z^{\ast} M_{-})|0\rangle,
 \end{eqnarray}
 with $z=\frac{\xi}{|\xi|}\tanh(\xi)$. Straightforward calculations leads one to the corresponding normalized Gilmore-Perelomov coherent state, i.e.,
 \begin{equation}\label{Gilmore}
   |z \rangle_{GP}=N_{GP}(|z|^{2})\sum_{n=0}^{\infty}
   \left[\frac{\Gamma(n+s+\frac{3}{2})}{\Gamma(n+1)\Gamma(s+\frac{3}{2})}\right]^{1/2}z^{n}|n\rangle,
 \end{equation}
 with the normalization factor
 \begin{equation}\label{normal Gilmore}
   N_{GP}(|z^{2}|)=\left(1-|z|^{2}\right)^{\frac{s}{2}+\frac{3}{4}}.
 \end{equation}
 Now, after finding out the two distinct classes of coherent states, we are ready to investigate some of the nonclassical properties which are of the most importance in the content of coherent states in quantum optics studies. But, before paying attention to this subject, we should investigate the resolution of the identity for the above introduced states. If we became successful, the introduced states can be truly named as coherent states.

\textit{Resolution of the identity}: Any coherent state $|z\rangle$ should satisfy the resolution of the identity which is defined as follows
 \begin{equation}\label{Resolution}
   \int d^{2}z w(|z|^{2})|z\rangle \langle z|=\hat{I}=\sum_{n=0}^{\infty}|n\rangle \langle n|
 \end{equation}
 where $w(|z|^{2})$ is a positive weight function that we search for obtaining it. For the first class of states which is introduced in this section,  substituting (\ref{Barut}) in (\ref{Resolution}) and using the change of variables $z=re^{i\theta}$ and $r^{2}=x$ lead us to the following integral relation
 \begin{equation}\label{Resolution BG}
   \int_{0}^{\infty}\tilde{w}_{BG}(x)x^{n}dx=\frac{\Gamma(n+1)\Gamma(n+s+\frac{3}{2})}{\Gamma(s+\frac{3}{2})},
 \end{equation}
 where we have used the abbreviation $\tilde{w}_{BG}(x)=\pi w_{BG}(x)N_{BG}^{2}(x)$. This is indeed an inverse moment problem which can be solved by the well-known methods such as Mellin transform technique \cite{klauder}.
   However, using the definition of Meijer's $G$-function, it follows that \cite{20}
 \begin{eqnarray}\label{38}
 \int^{\infty}_{0}dx\;x^{k-1} &G^{m,n}_{p,q}&\!\left(\begin{array}{ll}
               \hspace{-2mm} \beta x &\hspace{-2mm}\left| \begin{array}{l}
                             a_{1},\;...,\;a_{n},\;a_{n+1},\;...,\;a_{p} \\
                             b_{1},\;...,\;b_{m},\;b_{m+1},\;...,\;b_{q}
                           \end{array}\right.
              \end{array}\hspace{-2mm}\right)
 \\ \nonumber &=&\frac{1}{\beta^{k}}\frac{\prod^{m}_{j=1}\Gamma\left(b_{j}+k\right)\prod^{n}_{j=1}
 \Gamma\left(1-a_{j}-k\right)}{\prod^{q}_{j=m+1}\Gamma\left(1-b_{j}-k\right)
 \prod^{p}_{j=n+1}\Gamma\left(a_{j}+k\right)}.
 \end{eqnarray}
 Comparing equations (\ref{Resolution BG}) and (\ref{38}),
  it can  be easily checked that the required weight function which can satisfy the integral equation (\ref{Resolution})
  reads as:
 \begin{equation}\label{weight BG}
   w_{BG}(x)=\frac{_{0}F_{1}(s+\frac{3}{2},x^{2})}{\pi \Gamma(s+\frac{3}{2})}
   G_{2,4}^{4,0} \left( x \  \Bigg\vert \  {(0,s+1/2) \atop (0,0,s+1/2,s+1/2)} \right)
 \end{equation}
 where $G$ is the well-known Meijer function and  ${}_{0}F_{1}(a; x)$ is the generalized hypergeometric function.\\
About the Gilmore-Perelomov states, by substituting (\ref{Gilmore}) in (\ref{Resolution}) we will have:
 \begin{equation}\label{Resolution GP}
   \int_{0}^{1}\tilde{w}_{GP}(x)x^{n}dx=\frac{\Gamma(n+1)\Gamma(s+\frac{3}{2})}{\Gamma(n+s+\frac{3}{2})},
 \end{equation}
 where we have used $\tilde{w}_{GP}(x)=\pi w_{GP}(x)N_{GP}^{2}(x)$. Similarly, in this case it is easy to check that the required weight function
 has the following form
  \begin{equation}\label{weight GP}
   w_{GP}(x)=\frac{\Gamma(s+\frac{3}{2})}{\pi \Gamma(s+\frac{1}{2})}(1-x)^{-2} {}_{2}F_{1}\left(0,0;s+\frac{1}{2};1-\frac{1}{x}\right),
 \end{equation}
where ${}_{2}F_{1}(a_1,a_2;x)$ is the generalized hypergeometric function.

 A few words seem to be necessary about the limits of the integrals in  (\ref{Resolution BG}) and (\ref{Resolution GP}). Indeed, any quantum state belongs legally to a
 Hilbert space if its normalization constant satisfy the inequality $0 < N < \infty$. This led us to the unit disk and whole of the complex plane  respectively for Barut-Girardello  and Gilmore-Perelomov state.

 \section{Nonclassical features of the introduced states}\label{sec-n5}
  In this section, we briefly discuss some of the  criteria of nonclassicality which are usually  used in the relevant literature \cite{cpb7},
and will be helpful for investigating the nonclassicality exhibition of our introduced states in the continuation of the paper.
 To achieve this aim, we refer to the normal
 squeezing,  amplitude-squared squeezing and sub-Poissonian statistics as the most popular nonclassicality signs. To mention the common feature of the above criteria we should indicate that if a state is squeezed or its quantum statistics is sub-Poissonian in a region of allowed space, the corresponding Glauber-Sudarshan $P$-function will not be positive definite, i.e., it has no classical analogue. But, we would like to imply that finding this function is usually a hard task  to do. Altogether, each of the above effects is in fact
 a sufficient (not necessary) condition for a quantum state to belong to the family of nonclassical states.
\begin{itemize}
\item
  {\it Normal squeezing:} In order to examine the quantum fluctuations of the quadratures of the field \cite{walls} we introduce the Hermitian operators
 \begin{equation}\label{xp}
 X_1=\frac{M_{-}+M_{+}}{2},  \qquad P_1=\frac{M_{-}-M_{+}}{2i}.
 \end{equation}
 The squeezing parameters can be defined as follows
 \begin{equation}\label{squeeze1}
   S_{\gamma}=\frac{(\Delta\gamma)^{2}}{\sqrt{\frac{1}{4}|\langle [X_1, P_1]\rangle|^{2}}}-1,\;\; \gamma =X_1, P_1,
 \end{equation}
 where ${(\Delta\gamma)^{2}} = \langle \gamma ^2 \rangle - \langle \gamma \rangle ^2$.
Using (\ref{xp}) in (\ref{squeeze1}) we obtain the following first-order squeezing parameters:
\begin{eqnarray}\label{sx}
 S_{X_1}&=&\frac{1}{2 \langle M_0 \rangle}(\langle M_-^{2} \rangle+\langle M_-M_+ \rangle+\langle M_+M_- \rangle+\langle M_+^{2} \rangle \nonumber\\&-&\langle M_-^{2} \rangle-\langle M_+^{2} \rangle-\langle M_- \rangle \langle M_+ \rangle-\langle M_+ \rangle\langle M_- \rangle)-1
 \end{eqnarray}
 \begin{eqnarray}\label{sy}
  S_{P_1}&=&\frac{1}{2 \langle M_0 \rangle}(-\langle M_-^{2} \rangle+\langle M_-M_+ \rangle+\langle M_+M_- \rangle-\langle M_+^{2}
 \rangle \nonumber\\ &+& \langle M_-^{2} \rangle+\langle M_+^{2} \rangle-\langle M_- \rangle \langle M_+ \rangle-\langle M_+ \rangle\langle M_- \rangle)-1,
\end{eqnarray}
 where all expectation values can be straightforwardly obtained with respect to both classes of states.
 A state is squeezed in $X_1$ or $P_1$ if it satisfies the inequalities $-1\!<\!S_{X_1}\!<\!0$ or $-1\!<\!S_{P_1}\!<\!0$, respectively.
 \\
 By comparing the two displayed diagrams in figure (1) it can be realized that the squeezing of Gilmore-Perelomov coherent states occurs only in $P_1$ direction in all real space. Our calculations about the squeezing for Barut-Girardello coherent states  in $X_1$ and $P_1$, show that this parameter is approximately $10^{-15}$ ($\cong0$) for both operators. Thus, in view of squeezing property, it is deduced that the Barut-Girardello states associated with the considered system behave similar to the linear harmonic oscillator.
\item {\it Amplitude-squared squeezing:} Following the proposal of Hillery  \cite{hillery}, to evaluate this parameter we define the Hermitian operators
 \begin{equation}\label{XP}
 X_2=\frac{M_{-}^{2}+M_{+}^{2}}{2},  \qquad P_2=\frac{M_{-}^{2}-M_{+}^{2}}{2i}.
 \end{equation}
 The squeezing conditions in $X_2$ or $P_2$ are respectively given by $-1\!<\!S_{X_2}\!<\!0$ or $-1\!<\!S_{P_2}\!<\!0$, where $S_{X_2}$ and $S_{P_2}$ defined as follows:
 \begin{equation}\label{amplitude}
 S_{\ell}=\frac{(\Delta\ell)^{2}}{\sqrt{\frac{1}{4}|\langle [X_2, P_2]\rangle|^{2}}}-1,\;\; \ell=X_2, P_2
  \end{equation}
with the definition ${(\Delta \ell)^{2}} = \langle \ell ^2 \rangle - \langle \ell \rangle ^2$.
The explicit form of the latter parameters for second-order squeezing can also be obtained in a similar manner to what is done in equations (\ref{sx}) and (\ref{sy}) for the normal squeezing.
In figure 2 amplitude-squared squeezing has been plotted against real $z$
for Gilmore-Perelomov coherent states, in $X_2$ and $P_2$ direction. According to the diagrams have been placed in figure 2, it appears that, like the first-order squeezing, for this state again amplitude-squared squeezing  occurs only in $P_2$ direction. Also, this parameter for Barut-Girardello states in the two directions $X_2$ and $P_2$ takes approximately the value of zero. This latter observation is also the same as the situation which is likely occurred in ordinary coherent states.
 \item {\it Mandel parameter:} To establish the quantum statistical properties and as another nonclassicality sign, we calculate the Mandel parameter \cite{mandel}, which can be generalized as:
 \begin{equation}\label{mandel}
   Q=\frac{\langle M_{+}^{2}M_{-}^{2} \rangle - \langle M_{+}M_{-} \rangle^{2}}{\langle M_{+}M_{-} \rangle}-1.
 \end{equation}
 The state for which $Q\!=\!0$, $Q\!<\!0$ and $Q\!>\!0$ respectively corresponds to the Poissonian (standard coherent states), sub-Poissonian (nonclassical states) and super-Poissonian (classical states) statistics.
 In figures 3 Mandel parameter has been plotted versus real $z$ for Gilmore-Perelomov coherent states. As is shown, these states behave nonclassically features in a finite region of space. Our numerical calculations of this parameter for Barut-Girardello states takes the fixed value $Q=-1$.  Therefore, their quantum statistics is such that Mandel parameter gets the lowest possible negative value, like the number states of harmonic oscillator which sometimes have been called as the most nonclassical states.
\end{itemize}

 Adding the above observations, we may conclude that, from the point of view of squeezing properties the
 pseudoharmonic oscillator Barut-Girardello coherent state which is a particular superposition of number states (of pseudoharmonic oscillator) behaves like the canonical coherent states $|z\rangle=\exp(-|z|^2/2)$ $\sum_{n=0}^\infty \frac{z^n}{\sqrt{n!}}|n\rangle_{HO}$.
 However, taking into account the quantum statistical properties of the field, the
 pseudoharmonic oscillator Barut-Girardello coherent state behaves like the number states of harmonic oscillator
 (note that $|n\rangle_{HO} \neq |n\rangle$, where $|n\rangle$ in the present paper is related to the eigenkets of pseudoharmonic oscillator potential).
 So, in this view, the latter state can also be named as the most nonclassical states, too.

    \section{Summary and conclusion}\label{sec-n5}
In summary, in this paper we obtained the eigenvalues and eigenfunctions of one-dimensional pseudoharmonic oscillator  by the factorization method. The presented approach allows us to find out appropriate generalized creation and annihilation operators by which one can raise and lower arbitrary eigenkets of the considered system. Then, we have explored that the system under consideration possesses a symmetrical realization of a particular dynamical group; it is shown that the constructed ladder operators can constitute the generators of the  $su(1,1)$ Lie algebra. Next, we have produced the explicit form of Barut-Girardello and Gilmore-Perelomov coherent states for the considered system. The resolution of the identity, as the most important task for establishing any state as a class of "coherent states" is established analytically and so the over-completeness relation for both states are demonstrated. Finally, we could highlighted some of the nonclassical properties, such as first-order,  amplitude-squared squeezing and sub-Poissonian statistics for the introduced states, numerically.
Interestingly,  we have also found that, in particular, considering squeezing properties, the
pseudoharmonic oscillator Barut-Girardello coherent state behaves such as the canonical coherent states, however taking into account the quantum statistical properties, the
latter state behaves like the number states of harmonic oscillator.
Although, we have applied the method to pseudoharmonic oscillator in this paper, the approach can be used for other systems of interest, too.
As another example,  we have also done the same procedure to the well-known problem:
a quantum particle in an infinite square well potential, successfully, with the new and novelty results, which will be appeared in near future.

%

{\bf Figure Captions:}

Fig. 1  Plot of  squeezing parameters, $S_{X_1}$ (up diagram) and $S_{P_1}$ (down diagram) against real $z$ with $s=1$ for Gilmore-Perelomov coherent states.

Fig. 2  Plot of amplitude-squared squeezing parameters, $S_{X_2}$ (up diagram) and $S_{P_2}$ (down diagram) against real  $z$  with $s=1$
   for Gilmore-Perelomov coherent states.

Fig. 3   Plot of Mandel parameter with $s=1$  for Gilmore-Perelomov coherent states  versus real  $z$.

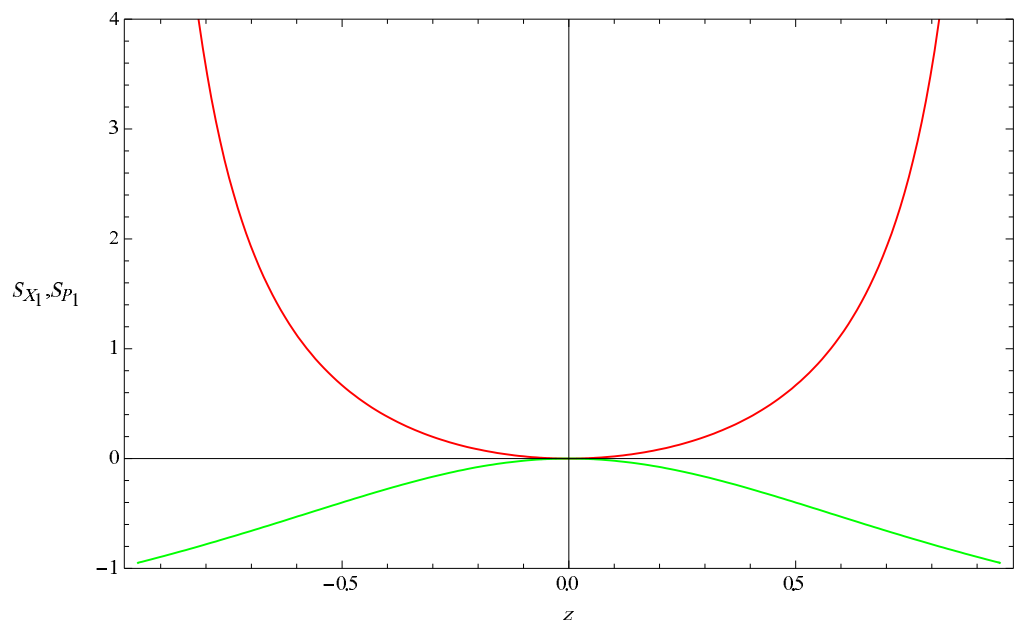
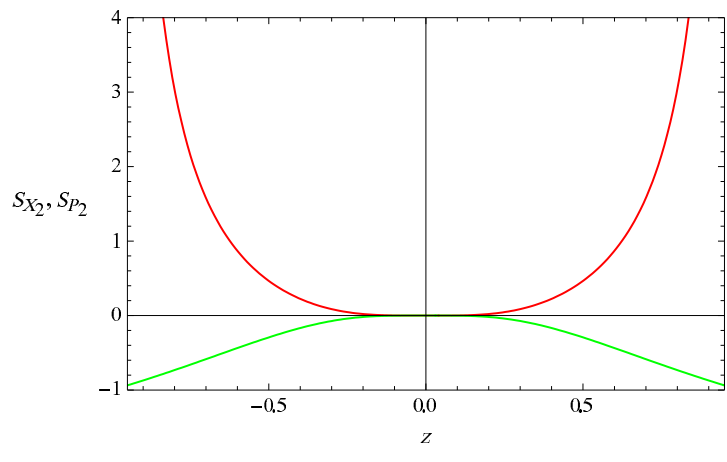
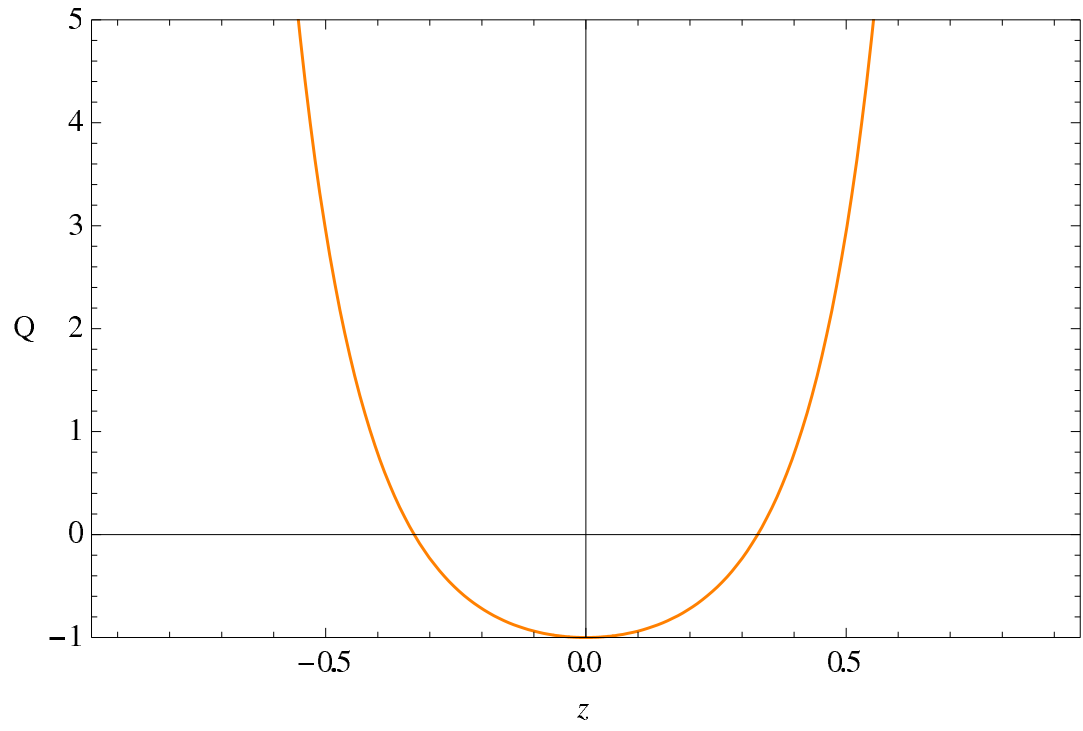
\end{document}